# Resistive-nanoindentation on gold: experiments and modeling of the electrical contact resistance

Fabien Volpi[a], Morgan Rusinowicz[a], Solène Comby-Dassonneville[a], Guillaume Parry[a], Chaymaa Boujrouf[a], Muriel Braccini[a], Didier Pellerin[b], Marc Verdier[a]

[a]Univ. Grenoble Alpes, CNRS, Grenoble INP, SIMaP, 38000 Grenoble, France

[b]Scientec/CSInstruments, 91940 Les Ulis, France

Corresponding author: fabien.volpi@grenoble-inp.fr

**Abstract**

This paper reports the experimental, analytical and numerical study of resistive-nanoindentation tests performed on gold samples (bulk and thin film). First the relevant contributions to electrical contact resistance are discussed and analytically described. A brief comparison of tests performed on gold and on natively-oxidized metals highlights the high reproducibility and the voltage-independence of experiments on gold (thanks to its oxide-free surface). Then the evolution of contact resistance during nanoindentation is fully explained in terms of electronic transport regimes: starting from tunneling, electronic transport is then driven by ballistic conduction before ending with pure diffusive conduction. The corresponding analytical expressions, as well as their validity domains, are determined and compared with experimental data, showing excellent agreement. From there, focus is made on the diffusive regime. Resistive-nanoindentation outputs are fully described by analytical and finite-element modeling. The developed numerical framework allows a better understanding of the main parameters: it first assesses the technique capabilities (validity domains, sensitivity to tip defect, sensitivity to rheology, effect of an oxide layer,…), but it also validates the different assumptions made on current line distribution. Finally it is shown that a simple calibration procedure allows a well-resolved monitoring of contact area during resistive-nanoindentation performed on samples with complex rheologies (ductile thin film on elastic substrate). Comparison to analytical and numerical approaches highlights the strength of resistive-nanoindentation for continuous area monitoring.

# 1. Introduction

The fine description and analysis of the contact between two solids is of both fundamental and technological interest [1,2]. From an academic point of view, the correlation between small-scale electron transport and mechanical loading can enrich the fundamental understanding of contact mechanics (effective contact area, small-scale mechanics,…). From a technological point of view, the improvement of engineered electrical contacts (switches, relays, microelectronic packaging,…) relies on the combination of finely controlled mechanical loads with in-situ monitoring of electrical conduction. In such a context, the electrical-functionalization of mechanics-oriented characterization techniques appears as a keystone able to control and monitor both mechanical and electrical properties of solids brought into contact.

Nanoindentation is a well-known technique dedicated to the local mechanical testing of materials at nanoscales (laterally and in-depth) [3,4]. In the last decade, numerous efforts have been made to expand the capabilities of this technique [5]: SEM imaging [6], high temperature testing [7], electrochemical nanoindentation [8], multi-field nanoindentation [9],… The coupling of indentation with electrical measurements was first initiated with large-scale indenters (macro- and micro-indentation) and finally extended to nanoindentation. This coupling was driven by a wide spectrum of motivations such as the local monitoring of phase transformation [10-16], the study of native oxide fractures [17-19], the characterization of piezoelectric materials [20,21], the characterization of particles dedicated to packaging [22], the investigation of MEMS operation [23,24], the monitoring of thin film dielectric behaviors [25,26] and the contact area computation during nanoindentation tests [27-32]. The latest point is of particular interest for the quantitative analysis of raw nanoindentation measurements since the contact area $A_c$ is the missing experimental data necessary for the extraction of both sample hardness and Young's modulus. Similar to standard indentation procedures (Vickers, Brinell) at the macroscopic scale, indirect methods to access to $A_c$ are based on post-mortem measurements of small size indent imprints by Atomic Force Microscope (AFM) or Scanning Electron Microscope (SEM). However, this approach only allows extracting hardness and elastic modulus at maximum load. Simple analytical models can be used to evaluate $A_c$ from experimental data (load, displacement and contact stiffness) on a semi-infinite body. A widely used approach relies on a partitioning of elastic/plastic displacement based on the pure elastic displacement solution (Oliver and Pharr method [33]). It can therefore only be used for materials showing sink-in behaviors (generally concerning materials with low stiffness-to-yield stress ratios). A major problem is that this method does not take into account pile-up phenomena [34]. An alternate approach proposed by Loubet [35] remains valid for both sink-in and pile-up contact behavior and relies on a phenomenological observation made on a wide range of materials. Nevertheless both methods are only valid for indenting a semi-infinite body and cannot apply for more complex boundary conditions with dissimilar material properties such as inclusion in a matrix or thin film on substrate. Regardless of these approaches, resistive-nanoindentation (sometimes referred as nano-ECR, for nanoindentation-electrical contact resistance) can be used to provide additional inputs to the quantitative analysis of indentation tests [31,36]. Thus this technique is expected to be an experimental alternative to standard analytical models, especially through the direct monitoring of contact area.

In terms of numerical modeling, even though the mechanics of nanoindentation have been extensively studied, the modeling of electrically-coupled nanoindentation is scarce and essentially focused on piezoelectric materials [37-39]. Finite-element modeling (FEM) of resistive-nanoindentation is essentially used to provide mechanics-related data that are post-processed to extract electrical resistance from analytical formulas [40-42]. Very few finite-element analyses of both electrical and mechanical behaviors have been reported [43-45].

The present paper reports a comprehensive study of resistive-nanoindentation experiments performed on oxide-free metals (focus is made on Au). It aims at giving a quantitative description of the physical processes that drive the electrical/mechanical behaviors of the contact during such experiments. The evolution of the electrical contact resistance is fully explained (from the earliest to the latest indentation stages) through analytical and numerical approaches. A step-by-step procedure is also set up in order to monitor the contact area during the indentation of a thin film, thus illustrating the strength of in-situ electrical/mechanical coupling. The developed modeling framework (analytical and numerical) is used to support the overall system description and to confirm the different assumptions made. A schematic of the overall methodology developed in the paper is given in Fig. 1.

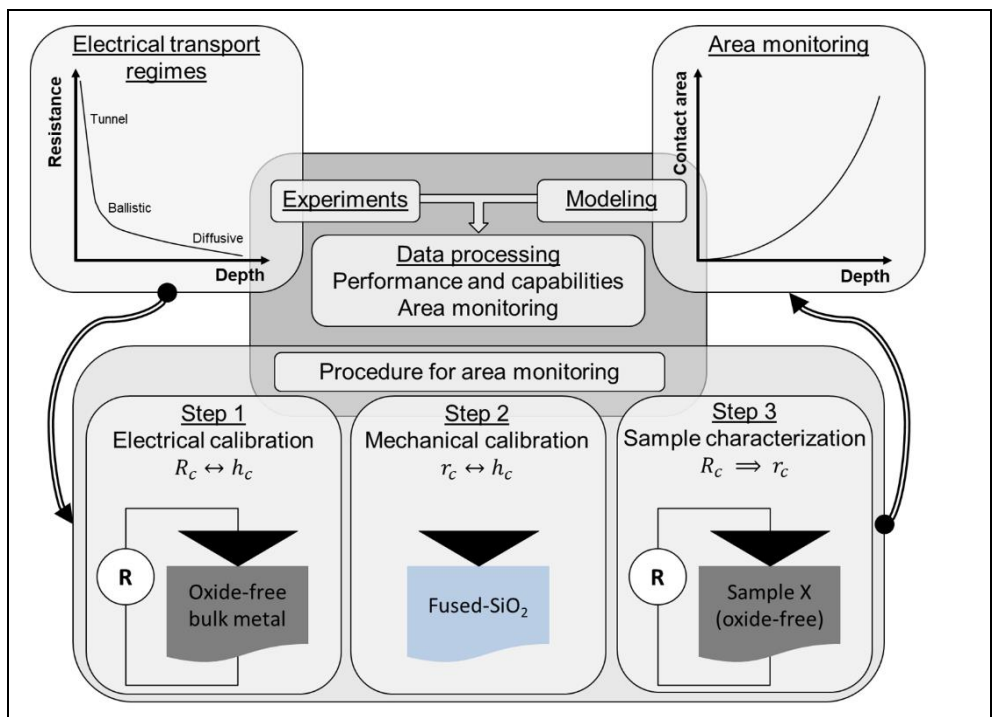

Figure 1: Schematic of the methodology developped in the present paper.

## 2. Experimental and numerical details

2.1. Resistive-nanoindentation set-up

Resistive-nanoindentation experiments were performed with a lab-made set-up, which combines different commercial devices (Fig. 2). The nanoindentation head is a commercial actuator (InForce 50 actuator from KLA-Nanomechanics Inc), with a dynamic displacement resolution better than 0.01 nm (capacitive gauge system). The indenting tip is fixed on a 3 mm-long ceramic extension, which is screwed on a 1.5 cm-long tungsten extender. The specimen is mounted on a double-side copper-coated epoxy plate. The ceramic extension and the epoxy plates are necessary to electrically insulate the conductive tip and the specimen from the grounded set-up frame. Electrical contacts (to the tip and the specimen) are made with thin copper wires connected to fixed sockets. Actuator and sample displacements are performed with linear positioners from SmarAct GmbH. Typical travel ranges are at the cm scale with a ~1 nm resolution. The overall stiffness of the instrument frame has been checked by indenting fused-silica specimen. Frame stiffness in the range of $10^6$ N/m has been extracted, thus validating the mechanical behavior of the overall set-up. During nanoindentation tests, the contact stiffness $S_c$ was measured continuously (Continuous Stiffness Measurement (CSM) mode) at an oscillating frequency of 100 Hz.

Resistance measurements were conducted with a ResiScope apparatus from Scientec [46]. This device (originally dedicated to Conductive-Atomic Force Microscopy measurements) is able to measure electrical resistance over 10 decades (from 100 Ω to 1 TΩ), with a refresh time circa 1 kHz. DC voltage is applied to the specimen while current is measured from the indentation tip with guarded connectors. Both electrical and mechanical responses are then acquired simultaneously at a typical frequency of 500Hz.

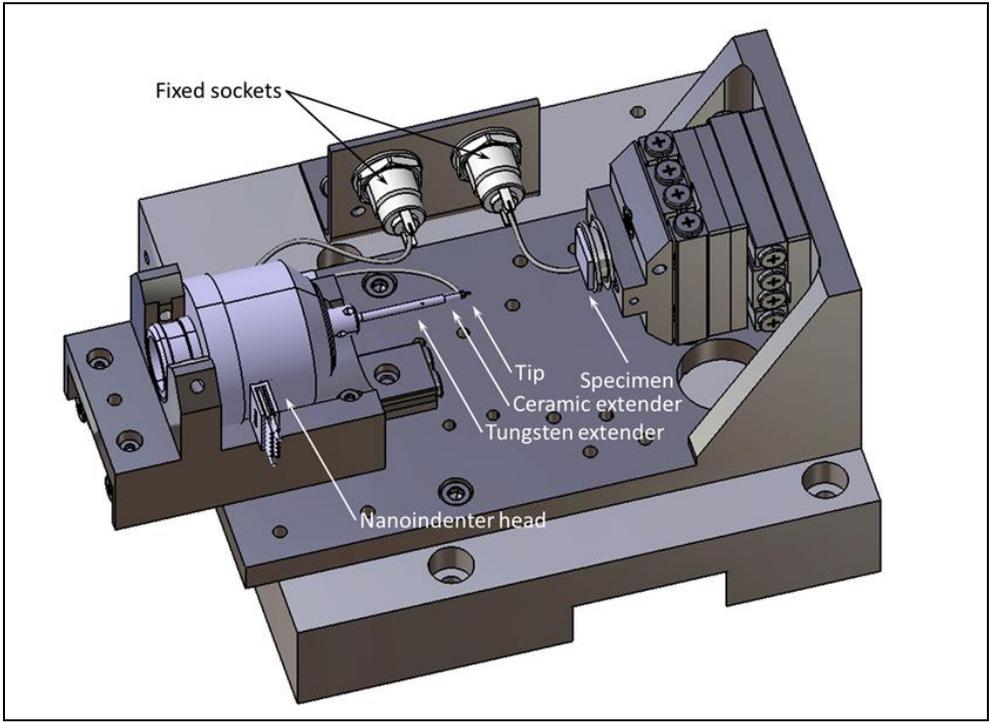

Figure 2: General view of the experimental set-up (CAD drawing).

## 2.2. Tips

Nanoindentation tests were performed with Berkovich boron-doped diamond (BDD) tips with resistivities in the range [0.2-2] Ω.cm from Synton-MDP. The analytical relationship between the tip cross-section $A_c$ and contact depth $h_c$ (the so called tip 'shape function') has been obtained by either of the two following methods: direct AFM imaging of the tip, and the calibration method described by Oliver and Pharr (indentation of fused $SiO_2$) [34]. For data processing, a second-order polynomial fitting of this relationship was used to relate analytically contact area to contact depth. The contact radius $r_c$ was then defined as follows:

$$r_c = \sqrt{\frac{A_c}{\pi}} \qquad (1)$$

Real tips display an unavoidable rounded apex, as shown in Fig. 3. This apex can be modelled as a sphere defined by its radius of curvature $r_{roc}$. The tip defect $h_0$ (height of the missing apex) was determined by extrapolating the self-similar domain of the tip 'shape function' to the depth axis. The defect extent $h^*$ (transition between spherical and self-similar shapes) was identified from the slope change on the tip 'shape function'.

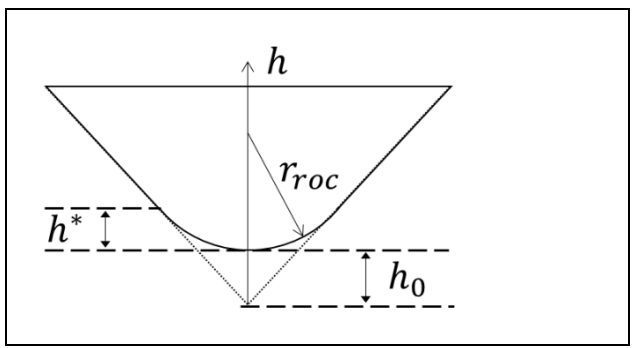

Figure 3: Schematic description of tip defect.

## 2.3. Samples

Three different bulk samples and a thin film were characterized. The bulk specimens considered are single crystals of Au (100-oriented), Cu (100-oriented) and Al (111-oriented) from MaTecK GmbH. The Au surface was prepared by fine polishing, leading to a 3.0 nm RMS roughness (measured by AFM). This polishing step led to hardening of the sample top-surface. The thickness of this hardened layer is at the micrometer-scale. Cu and Al surfaces were also electropolished after annealing and mechanical polishing. Native oxide thicknesses were measured by X-Ray Reflectometry and Angle-Resolved X-Ray Photoelectron Spectrometry: ~1 nm-thick on Cu and 6-9 nm-thick on Al. The thin film was a 200 nm-thick polycrystalline Au film deposited by evaporation on a sapphire substrate (~2.5 nm RMS roughness measured by AFM).

## 2.4. Finite-element modeling

Numerical calculations are carried out using finite-elements modeling with the ABAQUS software [47]. The indenter and the substrate are modeled as deformable bodies coming into contact. A potential bias is applied between the top surface of the tip and the bottom surface of the sample, while the indenter is first pushed down into the substrate up to the target depth, and then withdrawn. A coupled mechanical-electrical procedure is used, in order to capture both the deformation and the distribution of currents in the system. The indenter is modeled as an elastic linear body, while the substrate is elastic-plastic. The electrical conduction is assumed to follow a pure Ohmic law. As the current flow between the substrate and the tip is crossing the indenter/substrate contact surface, it is crucial to obtain a good prediction of this contact during loading – it is actually a key point motivating this numerical modeling. For this purpose, a classical "hard" mechanical contact model is chosen between the indenter and substrate surfaces (minimizing the penetration of the "slave" surface into the "master" surface). An electrical contact property is also chosen, assuming very high conductivity of the indenter/substrate interface at contacting nodes. An implicit scheme is used to solve the coupled problem. Due to the high nonlinearity of the mechanical problem (plasticity, contact, large deformations), the mechanical loading is applied incrementally. The steady state electrical conductivity equations are solved at the same time as the mechanical equilibrium equations. Finally, a post-processing step is carried out after the FEM calculations are done, in order to compute the contact area, the electrical resistance of the system, and the reaction force of the substrate on the indenter during loading/unloading. All the details of the FE modeling are provided in appendix.

## 3. Preliminary considerations

### 3.1. Importance of contact area in nanoindentation – Effect of material rheology

One of the main interests of nanoindentation is the measurement of hardness and Young's modulus, which are extracted from two mechanical outputs (the applied load $L$ and the contact stiffness $S_c$) combined with the contact area $A_c$ according to the following equations:

Sample hardness $H$ is obtained from Eq. 2:

$$H = \frac{L}{A_c} \qquad (2)$$

Sample Young's modulus is determined thanks to Sneddon's relation [48]:

$$E^* = \frac{\sqrt{\pi}}{2} \frac{S_c}{\sqrt{A_c}} \qquad (3)$$

with $E^*$ the reduced modulus, expressed as:

$$E^* = \left(\frac{1-\nu_{tip}^2}{E_{tip}} + \frac{1-\nu_{sample}^2}{E_{sample}}\right)^{-1} \qquad (4)$$

with $\nu_{tip}$, $\nu_{sample}$, $E_{tip}$, $E_{sample}$ the tip and sample Poisson's ratios and moduli, respectively.

These equations underline the need to determine the contact area $A_c$ independently of any mechanical outputs. $A_c$ depends both on the tip geometry and on the contact depth $h_c$ ($h_c$ being the height of the tip effectively in contact with the sample (Fig. 4)). This contact depth can be either greater or smaller than the penetration depth $h$ ($h$ being the depth reached by the tip from the initial specimen surface), which is the magnitude monitored experimentally. The tip 'shape function' can be unambiguously determined by direct AFM imaging of the tip or by calibration on a reference sample [47]. As shown in Fig. 4, the contact depth $h_c$ depends strongly on sample rheology: materials with low (large) stiffness-to-yield stress ratio display sink-in (pile-up) profiles around the tip. When pile-up is large, standard model underestimates $A_c$ by as much as 60% [49], thus introducing a 60% error on the measured hardness and 30% error on Young's modulus. This point highlights the need for an alternative method to extract the true contact area.

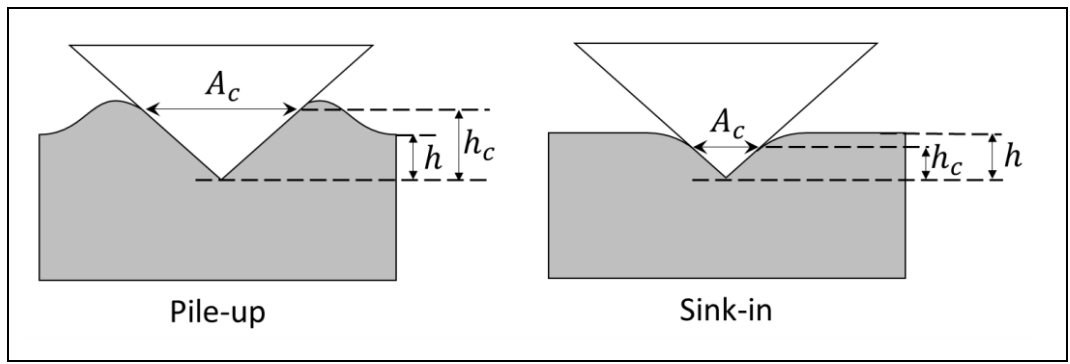

Figure 4: Effect of material rheology on contact area $A_c$ and contact depth $h_c$.

3.2. Individual contributions to the electrical contact resistance

The quantitative analysis of resistive-nanoindentation outputs first requires the identification of the individual contributions to the total measured resistance. As long as the electronic transport remains in diffusive regime (discussed in Section 4.2), the measured contact resistance is the sum of the following contributions (from left to right in Fig. 5):

$$R_{Contact} = R_{Set-up} + R_{Tip} + R_{Interface} + R_{Sample} \qquad (5)$$

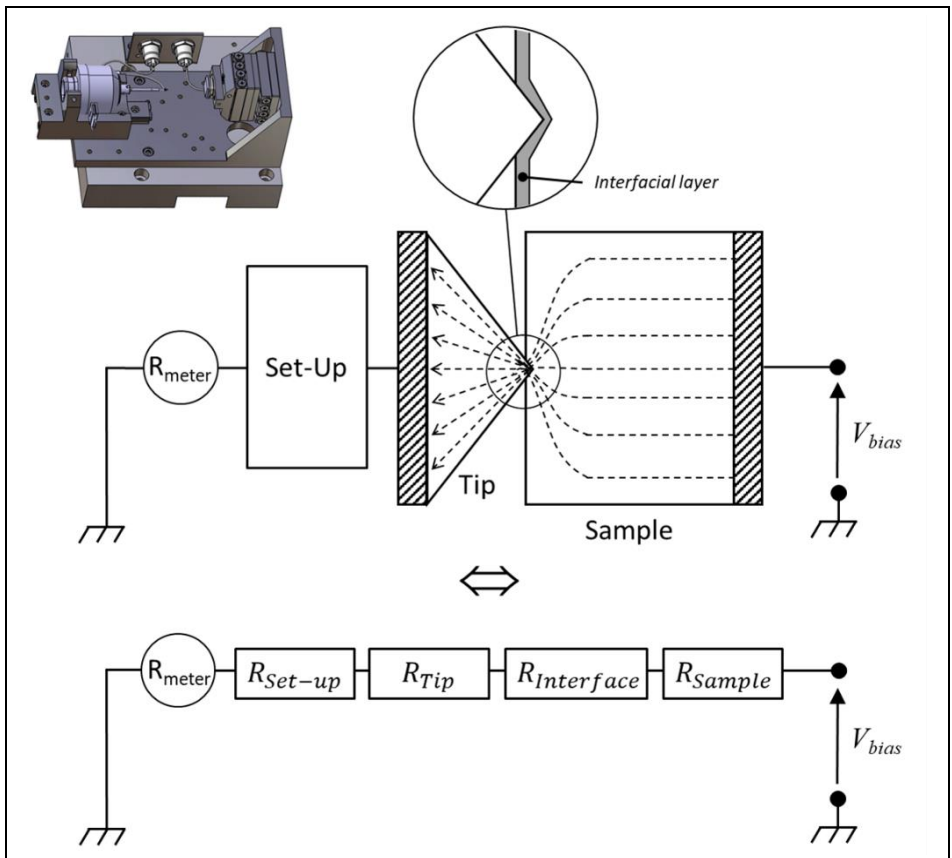

Figure 5: Schematic of the individual contributions to the measured resistance and the corresponding electrical cicruit. Dotted lines represent the current lines.

- A fixed resistance ($R_{Set-up}$) coming from the experimental set-up (wires, connectors, tip brazing,…). This resistance is an empirical characteristic of the set-up.

- The electrical resistance of the indentation tip ($R_{Tip}$). This resistance is proportional to the tip resistivity ($\rho_{tip}$). In the specific case of self-similar tips (Berkovich, cube-corner and conical geometries), simple geometrical considerations show that this resistance evolves as the reciprocal of the contact radius $r_c$ (Eq. 6) (adapted from [50]).

$$R_{Tip} \propto \frac{\rho_{Tip}}{r_c} \tag{6}$$

This expression relies on two strong assumptions: 1/ Above the contact level, the current distribution within the tip is homogeneous and homothetic along the tip axis and 2/ Below the contact level, the part of the tip that has already penetrated into the sample affects only linearly the tip resistance.

Regarding the first assumption, it has been shown by numerical modeling that the analytical 'stacked discs' approach that leads to Eq. 6 applies if corrected by a geometrical coefficient [51,52]. The second assumption is supported by the current line distribution within the tip: since the BDD tip is five orders of magnitude more resistive than metals, current lines are strongly localized at the periphery of the contact instead of being homogeneously distributed (Fig. 6 a). Nakamura et al [53] have shown that the electrical resistance of ring contacts is equivalent to the resistance of the corresponding full-surface contact divided by a shape factor $SF$. This shape factor describes the amount of current that flows through a peripheral ring of width $t_r$ normalized to the total current (Fig. 6 b). For now, we assume that this shape factor $SF$ is constant throughout indentation, which thus maintains the validity of Eq. 6. This assumption will be further confirmed by numerical modeling.

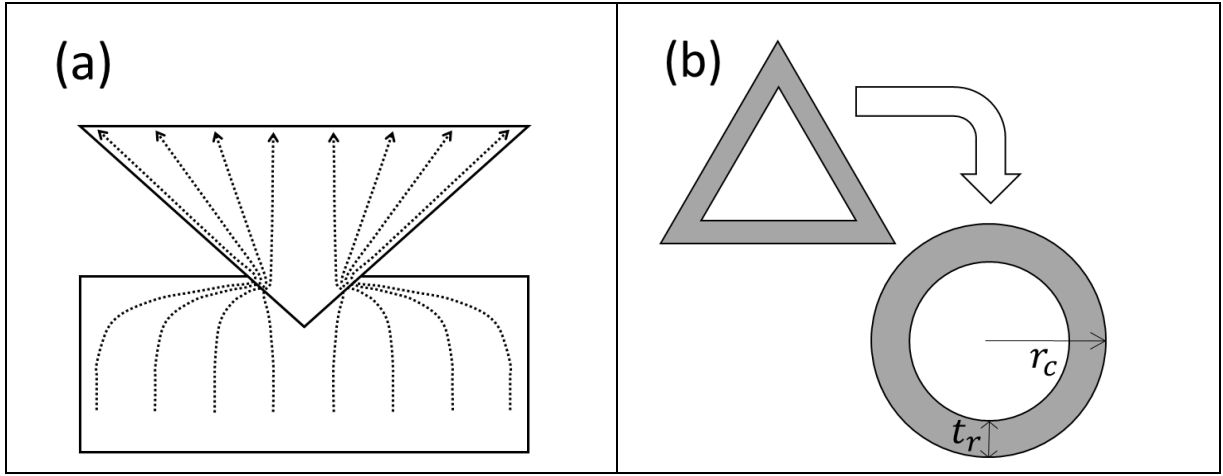

Figure 6: Schematic of current line distributions. (a) Vertical cross-section of current constriction at contact periphery. (b) Horizontal cross-section : conversion of the actual triangular cross-section into a circular ring-shaped cross-section (suitable for axisymmetric numerical modeling).

- The interfacial resistance ($R_{Interface}$), coming from a possible interfacial layer (oxide or organic contaminant). Depending on the specimen surface state, $R_{Interface}$ can either be neglected (Au case) or should be considered if the metal is oxidized (Cu and Al case). This resistance evolves as the square of the contact radius reciprocal [2]:

$$R_{Interface} \propto \frac{\rho_{Interface}}{r_c^2} \qquad (7)$$

- The spreading resistance within the sample ($R_{Sample}$), due to the current line constriction to the contact. This resistance is proportional to the sample resistivity. As resistivity of standard metals is 5 orders of magnitude lower than the BDD-tip resistivity, this spreading resistance can be safely neglected for the upcoming analysis. This assumption means that the resistive-nanoindentation technique is insensitive to material resistivity with BDD-tips. In the case of highly-conductive tips, $R_{Sample}$ should be considered, as discussed in [29,30].

Finally the overall measured contact resistance simply consists of a linear-quadratic relationship with the reciprocal of contact radius:

$$R_{Contact} = A + \frac{B}{r_c} + \frac{C}{r_c^2} \qquad (8)$$

with A and B two constants that depend only on the experimental set-up, and C a constant that describes the interfacial behavior. In the case of oxide-free samples, C is nil. Finally, the contact radius $r_c$ is the only magnitude related to the material. $r_c$ is the signature of material rheology (Fig. 4).

## 4. Results and discussion

### 4.1. Resistive-nanoindentation: general observations

A typical set of resistance-vs-penetration depth (R-h) curves obtained on bulk Au for different polarization biases is given in Fig. 7 a. As expected, the main trend is the overall decrease of resistance as the mechanical contact area increases [29]. This figure also illustrates the high reproducibility between all the tests. It also shows that the R-h curves are voltage-independent, which is the evidence of an ohmic-like electrical contact. This later point is supported by the linearity of current-voltage (I-V) curves recorded at different penetration depths (Fig. 7 b), in agreement with [29,30] for the indentation of oxide-free metals with highly conductive tips. This linearity is an essential property in order to use electrical signals for mechanical characterizations [54]. This ohmic behavior is consistent with the absence of any interfacial layer (oxide or organic contaminant) between tip and specimen. Thus the measured resistance given by Eq. 5 simply reduces to the sum of the tip resistance with the constant set-up resistance:

$$R_{Contact} = R_{Set-up} + R_{Tip} \qquad (9)$$

These experiments have also been performed under vacuum ($10^{-2}$ Torr) and led to identical results. These highly-reproducible and linear electrical characteristics make possible the fine analytical processing of resistive-nanoindentation data.

In order to test the influence of a surface oxide layer, resistive-nanoindentation experiments were also performed on natively-oxidized Cu and Al. These metals have been chosen as their native oxides display different conduction mechanisms: mixed (electronic/ionic) and ionic-only conduction, respectively. Fig. 7 c-d report R-h curves obtained on Cu and Al, respectively. A strong bias-dependence is observed on both samples, in accordance with the non-linearity of I-V characteristics of dielectric stacks. In addition, the noisy and dispersed evolution of the Cu and Al curves (essentially at low bias) is the signature of competing mechanisms like oxide cracking and electrochemical processes (anodic oxidation). The analysis of these data requires statistical considerations as well as the kinetic description of electrochemical-processes. A more refined description of natively-oxidized metals will be given elsewhere.

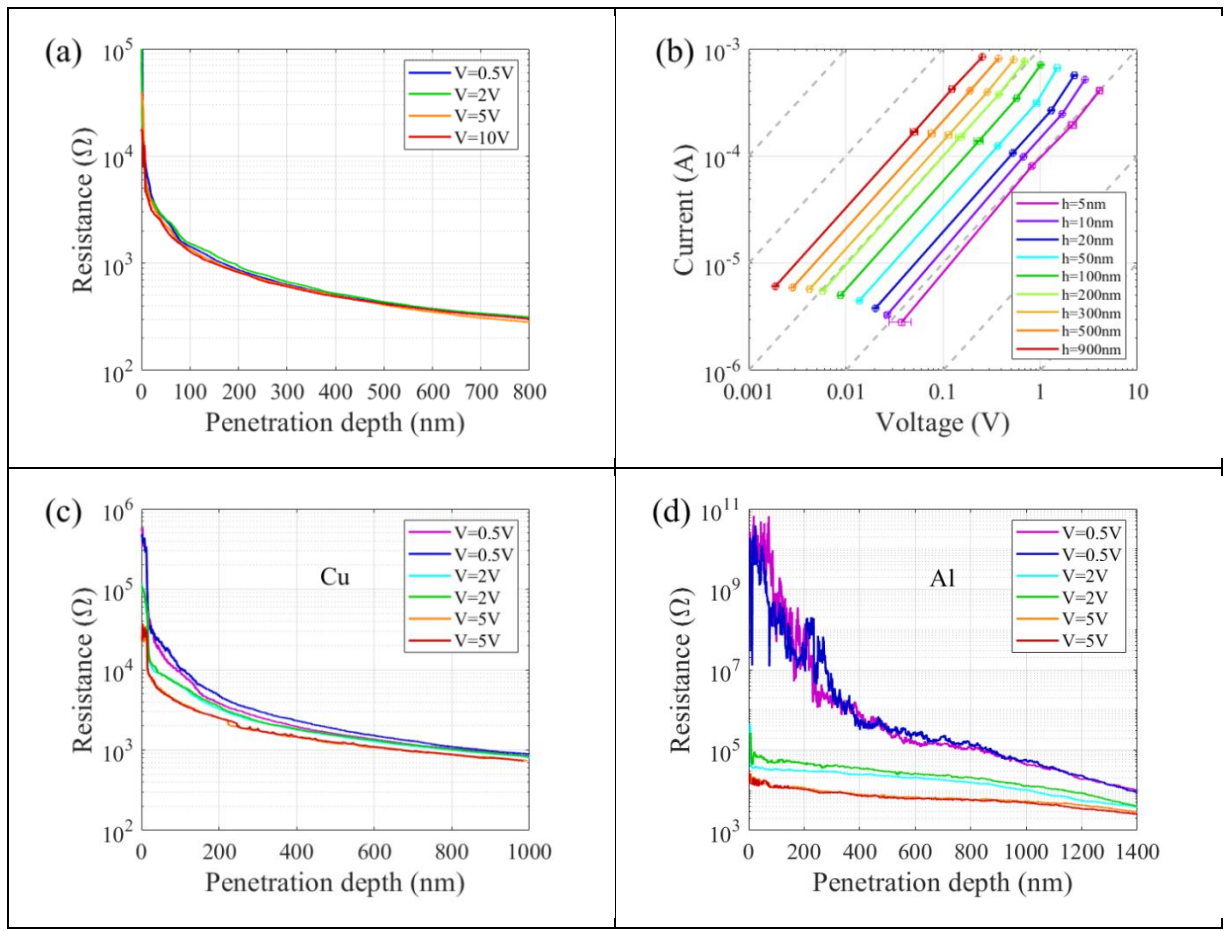

Figure 7: Resistive-nanoindentation results. (a) Set of R-h curves obtained on bulk Au. (b) Corresponding I-V curves for different penetration depths. (c) Set of R-h curves obtained on natively-oxidised bulk Cu. (d) Set of R-h curves obtained on natively-oxidised bulk Al.

## 4.2. Identification of electrical transport regimes

### 4.2.1. Experimental results

In order to discriminate the different conduction mechanisms that prevail at each indentation stage, a log-log plot of several resistive-nanoindentation tests is presented in Fig. 8 a. For these experiments, acquisition rate was set to 500 Hz (5 times larger than the CSM frequency). Before permanent contact is established, the tip displays white noise vibrations that lead to intermittent contacts, even before the surface is mechanically detected. Consequently, data recorded before surface detection (i.e. for negative penetration depth) cannot be plotted in log scale. For the sake of graphical visualization, the plot was artificially shifted along the X-axis by 2 nm, thus leading to the 'Shifted penetration depth' axis in Fig. 8 a. The corresponding effective penetration depths $h$ are given on the secondary X-axis.

Two main trends can be first observed. At the early stages (for effective penetration depths lower than ~1 nm) resistance is highly dispersed and drops rapidly over 7 orders of magnitude: from ~$10^{12}$ Ohms (the sensitivity limit of the ResiScope) down to ~$10^5$ Ohms. Beyond 1 nm, all curves tend to merge and to superimpose, in accordance with the high reproducibility observed in Fig. 7 a. The objective is now to describe this plot in terms of electrical transport regimes.

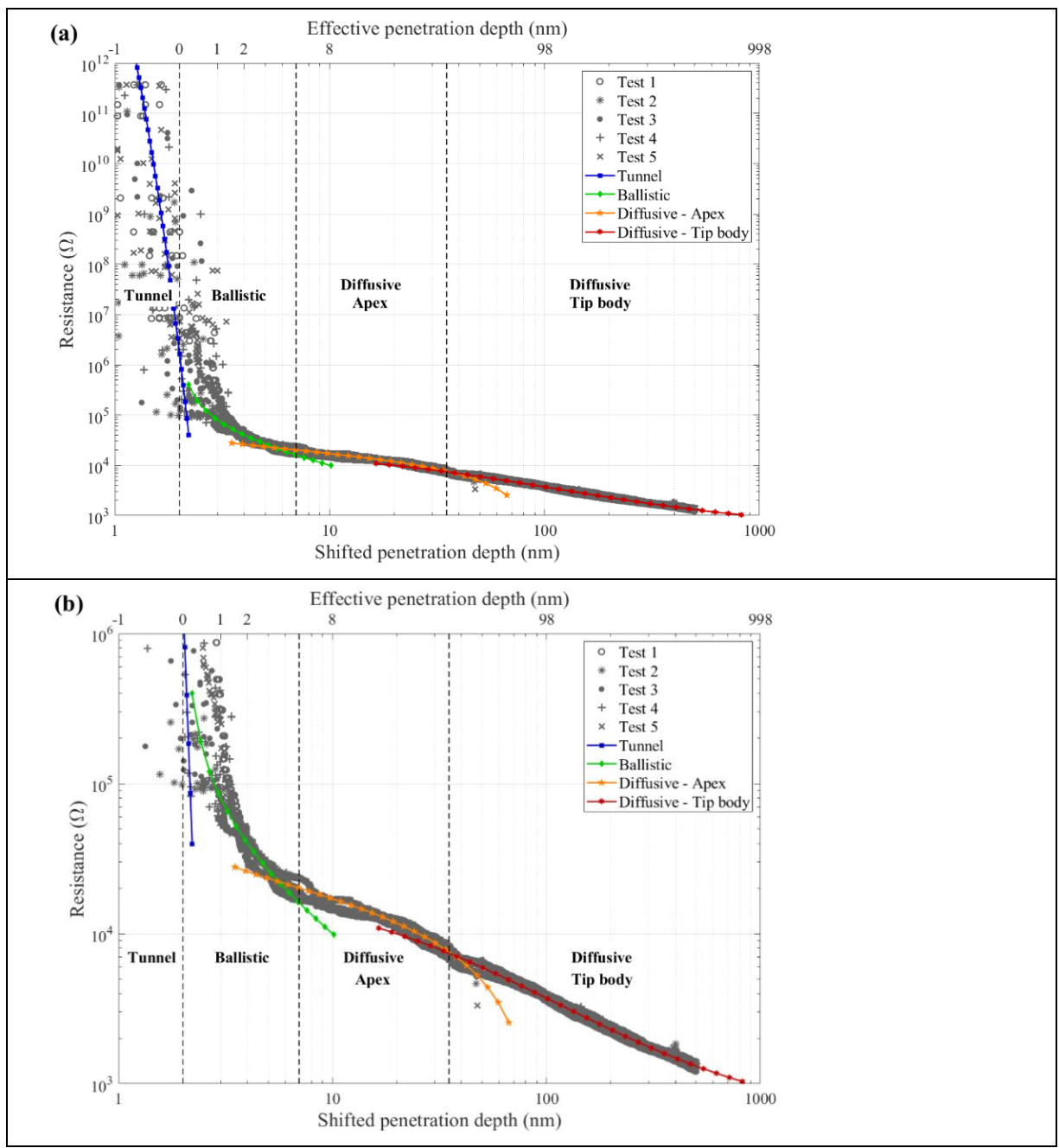

Figure 8: Resistive-nanoindentation tests on bulk Au in log-log scale (Applied bias = 2 V). (a) Full scale graph. (b) Zoom-in of (a) to highlight the mechanisms other than tunneling.

### 4.2.2. Tunneling regime

Before direct solid-solid contact is established, electrons can flow from tip to sample by tunneling effect. The corresponding tunneling resistance depends exponentially on the distance between surfaces:

$$R_{Tunnel} = A_T \, exp(-2 \, \alpha \, h_{QS}) \quad for \; h_{QS} \leq 0 \qquad (10)$$

with $A_T$ a constant, $h_{QS}$ the quasi-static penetration depth, $\alpha = \sqrt{2 \, m \, E_{Tunnel}}/\hbar$ and $E_{Tunnel}$ the height of the energy barrier to cross. The quasi-static penetration depth $h_{QS}$ is given by:

$$h_{QS} = h - \sqrt{2} \, \delta h \qquad (11)$$

with $\delta h$ the RMS tip oscillation (experimentally measured at 0.65 nm). A reasonable estimate of $E_{Tunnel}$ (3 eV) can be obtained by subtracting the applied bias (2 V) and an average of the negative electron affinity of diamond (~0.5 eV) [55] to the diamond band gap (5.5 eV). Only the pre-exponential factor $A_T$ remains an adjustable parameter for graphical representation. In this depth range, $h_c = h$.

### 4.2.3. Ballistic regime

The first solid-solid contacts are due to surface roughness: each discrete spot where the diamond tip squeezes local asperity becomes the locus for ballistic conduction (with a local contact radius $r_c$). On these highly localized spots, where $r_c$ is small compared to the electron mean free path ($\lambda_{mfp}$), electrons mainly collide elastically while crossing the contact interface. The corresponding ballistic resistance ($R_{Ballistic}$) evolves as the square of the contact radius reciprocal (Eq. 12) [2,30,56,57].

$$R_{Ballistic} \propto \frac{\lambda_{mfp}}{r_c^2} \qquad (12)$$

At such a shallow depth, the tip apex is inevitably rounded and can be modelled as a sphere with large radius of curvature $r_{roc}$ (estimated circa 150 nm for this tip, which is typical for used Berkovich tips). Simple geometrical considerations show that a depth $h_c$ leads to a contact radius $r_c$ given by Eq. 13:

$$r_c = \sqrt{2r_{roc}h_c - h_c^2} \tag{13}$$

As $r_{roc}$ is large compared to $h_c$, Eq. 12 simply drops to:

$$R_{Ballistic} = \frac{A_B}{h_c} \tag{14}$$

with $A_B$ a constant.

The transition to the next regime occurs when the contact radius $r_c$ equals the mean free path $\lambda_{mfp}$. In the present case, conduction in boron-doped diamond tip is provided by holes, with a mean free path in the 1-10 nm range [58,59]. In gold, the electron mean free path is 38 nm [2], which therefore becomes the relevant threshold. Solving Eq. 13 with $h_c$ as the unknown and replacing $r_c$ by $\lambda_{mfp}$ leads to a transition at:

$$h_B \cong \frac{\lambda_{mfp}^2}{2r_{roc}} \tag{15}$$

Numerical application shows that ballistic regime extends up to 5.0 nm, which exceeds the sample roughness (3.0 nm).

4.2.4. Diffusive regime

When the local contact radius is larger than the electron mean free path, electron flow can be safely considered as continuous through the averaging of electron collisions into a collective approach. This is the diffusive regime. Within this regime, classical electromagnetism shows that the constriction resistance evolves linearly with the reciprocal of the contact resistance (Eq. 6). As the tip shape evolves continuously from a spherical apex to self-similarity, two domains have to be distinguished.

The constriction resistance of the spherical tip apex can be deduced from simple geometrical considerations:

$$R_{Apex} = \frac{\rho_{Tip}}{\pi . r_{roc}} \times \left[ atanh\left(\frac{r_{roc} - h_c}{r_{roc}}\right) - atanh\left(\frac{r_{roc} - h^*}{r_{roc}}\right) \right] \qquad (16)$$

This expression applies as long as the contact depth is smaller than the defect extent $h^*$.

Beyond this point, self-similarity prevails, meaning that contact radius $r_c$ should be proportional to contact depth $h_c$. However, because of the missing tip defect $h_0$ (Fig. 3), the proportionality to $r_c$ only applies with the sum $h_c + h_0$ (referred as 'corrected contact depth'). Finally Eq. 6 turns into Eq. 17 to describe the main body of the tip:

$$R_{Tip\ Body} \propto \frac{\rho_{Tip}}{h_c + h_0} \qquad (17)$$

This latter expression is the cornerstone equation for the upcoming quantitative analysis of resistive-nanoindentation tests (Section 4.3).

4.2.5. Comparison to experimental data

The four transport mechanisms described by Eqs. 10, 16, 18 and 19 are plotted in Fig. 8 a, as well as their validity domains. Table 1 summarizes the characteristics of the conduction regimes: expression, upper bounds and adjustable parameters (all other constants were obtained by relevant regressions on experimental data). As it can be seen in Fig. 8 a, these analytical curves show excellent agreement with experimental data, despite the low degree of freedom on adjustable parameters. As tunneling mechanism tends to overwhelm other mechanisms (6 orders of magnitude against 3), a zoom-in of Fig. 8 a in the $10^3 - 10^6$ range is shown in Fig. 8 b to highlight the agreement of analytical models to experimental data in this range. It can be concluded that the proposed timeline of conduction mechanisms fully describes the evolution of resistance from the earliest stages down to the deepest indentation.

| Mechanism / Tip part | Equation | $h_c$ upper bound | Adjustable parameters |
|---|---|---|---|
| Tunneling | $R_{Tunnel} = A_T \exp(-2\,\alpha\,h_{QS})$ | 0 | $A_T = 1 \times 10^{14}$ |
| Ballistic | $R_{Ballistic} = \dfrac{A_B}{h_c}$ | $h_B \cong \dfrac{\lambda_{mfp}^2}{2 r_{roc}}$ | $A_B = 8 \times 10^4$ |
| Diffusive / Spherical tip apex | $R_{Apex} = \dfrac{\rho_{Tip}}{\pi \cdot r_{roc}} \times \left[ \operatorname{atanh}\left(\dfrac{r_{roc} - h_c}{r_{roc}}\right) - \operatorname{atanh}\left(\dfrac{r_{roc} - h^*}{r_{roc}}\right) \right]$ | $h^*$ | None |
| Diffusive / Tip body | $R_{Tip\,Body} \propto \dfrac{\rho_{Tip}}{h_c + h_0}$ | Maximum penetration depth | None |

Table 1: Electrical transport characteristics.

### 4.3. Data processing and numerical modeling of resistive-nanoindentation tests on bulk Au

Now the electrical transport regimes being identified throughout the indentation range, focus is made on the diffusive regime, which is the useful range to engineer nanoindentation outputs.

#### 4.3.1. Analytical data processing

As already shown (Fig. 7 a), raw resistive-nanoindentation outputs relate contact resistance $R_{Contact}$ to penetration depth $h$ ($R \leftrightarrow h$ relationship). Eq. 17 proposes to relate analytically contact resistance to contact depth $h_c$ ($R \leftrightarrow h_c$ relationship). This latter correspondence can be obtained experimentally by two independent methods: post-mortem AFM imaging of the residual imprints (combined to the tip 'shape function' ($r_c \leftrightarrow h_c$ relationship)) or the application of the standard Oliver and Pharr model [33] ($h \leftrightarrow h_c$ relationship). In the present case, because of its hardened surface, the Au specimen presents a sink-in profile, thus allowing the application of the Oliver and Pharr model. Consequently both methods were used and led to identical results. From there, the contact resistance can be plotted against the reciprocal of the 'corrected contact depth' (Fig. 9 a) and the contact radius (Fig. 9 b). These two linear plots confirm experimentally the linear dependence expected by Eq. 6 and 19.

The linear fitting of the $R_{Contact}$ vs $1/(h_c + h_0)$ plot (Fig. 9 a) allows the determination of the A and B regression constants in Eq. 8:

$$R_{Contact} = 245 + \frac{2.52 \times 10^5}{h_c + h_0} \qquad (18)$$

This calibration step establishes a one-to-one analytical relationship between contact resistance $R_{Contact}$ and contact depth $h_c$. As this calibration fully defines the experimental set-up (tip geometry and resistivity, series resistance,…), it allows the investigation of any sample as long as its surface is oxide-free and its resistivity is much lower than the tip resistivity. This linear dependence with the reciprocal of contact radius is the signature of an ideal tip-to-sample interface. Non-linearity can either rise from tip inhomogeneities (deviation from self-similarity, doping gradients,…) or from a nonlinear electrical behavior of the interface (rectifying contact, insulating layer,…). For instance, a quadratic dependence [28,31,54] is the signature of an interface-driven resistance, suggesting a nonlinear behavior of the contact (because of an oxide layer or a lowly doped diamond tip, as discussed in [54]). Non-linear behavior poses significant challenges for using electrical resistance for mechanical characterization: as the interface behavior depends on the nature of the indented material, a calibration-based procedure cannot be set. When linearity is observed, a universal contact area monitoring procedure can be trustfully applied for the characterization of any material displaying this linear behavior (oxide-free or equivalent).

It should also be noted that, if the material rheology is not adequate (pile-up profile), the Oliver and Pharr model should be avoided for this calibration step. In that case AFM imaging is necessary but might introduce slightly lousier fitting.

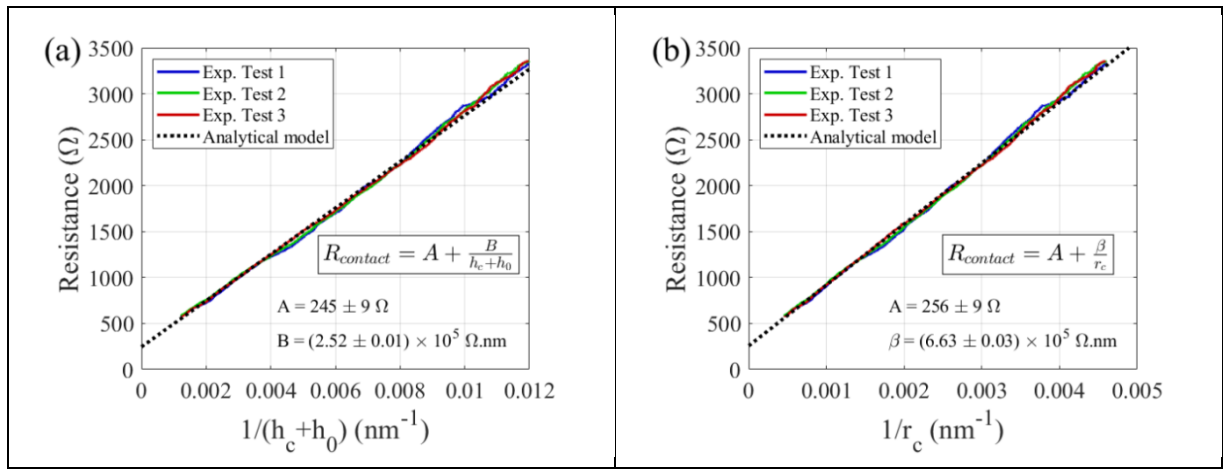

Figure 9: Plotting of the contact resistance against (a) the 'corrected contact depth' and (b) the contact radius.

### 4.3.2. Numerical modeling

Resistive-nanoindentation experiments were numerically modelled in order to identify its capabilities (bounds, sensitivity to tip defect, sensitivity to rheology, effect of an oxide layer,…) and to validate the assumptions made for the analytical model.

#### 4.3.2.1. Assessment of resistive-nanoindentation capabilities

Fig. 10 a-b compare the load and resistance curves obtained by numerical modeling to a set of experimental curves. In this system only few parameters are adjustable: magnitudes are either extracted experimentally (sample and tip resistivity, tip geometry…) or widely accepted in literature (constitutive law for Au with standard figures, polishing-induced surface hardening,…). Despite this low degree of freedom, an excellent agreement is obtained between experimental and simulated data, thus opening the door to fine analysis.

First the evolution of contact resistance with $1/(h_c + h_0)$ (Eq. 17) can be plotted and compared to experimental data (Fig. 10 c). In accordance with Fig. 9, the agreement is excellent, thus numerically validating the linear dependence proposed in Eq. 17. In addition, this numerical approach allows identifying the theoretical validity domain of this equation. Fig. 10 d illustrates the deviation from linearity on different simulated tips with varying defect extents ($h^* = 20$, 40 and 80 nm, see Fig. 3). As expected, deviation only occurs at the beginning of indentation, i.e. while indenting with the rounded tip apex. For the three simulated tips, deviations are lower than 1% for contact depth larger than $h^*$, which confirms $h^*$ as the lower limit of Eq. 17 proposed in Section 4.2.4. It also means that resistive-nanoindentation tests performed at shallow depth require lowly worn tips to be analyzed in a quantitative manner.

In order to test the sensitivity to material rheology, two boundary cases have also been tested: resistance curves obtained either on a realistic ductile specimen (leading to a pile-up profile) or on a hardened specimen (leading to a sink-in profile) are shown in Fig. 10 e. A ~20% difference is observed in terms of electrical resistance (for a 65% difference in contact area, in reasonable accordance with [49]). This sensitivity can be improved by reducing the set-up resistance: a 49 Ohm set-up resistance (instead of 245 Ohm) brings the sensitivity up to 30% (not shown). Once compared to the lowly-dispersed experimental curves (Fig. 9 for instance), this 20-30% resistance difference highlights the great sensitivity of resistive-nanoindentation to material rheology.

The case of natively-oxidized metals was also considered. Fig. 10 f compares the contact resistance evolution for a bulk metal covered or not by a 10 nm-thick highly-resistive film. First the quadratic dependence proposed in Eq. 7 is clearly identified. Depending on their chemistry and conduction type, real oxide resistivities cover several decades (typ. $10^{-3} - 10^{15}$ Ohm.cm). Despite the fact that simulated resistivity value (100 Ohm.cm) belongs to the lower range of this spectrum, the oxide resistance is one order of magnitude larger than the tip resistance. Consequently, in real cases, the oxide resistance is expected either to dominate all other signals (as shown in Fig. 7 d with native alumina) or to remain in a comparable range allowing fine analysis (copper oxide case in Fig. 7 c). In the case of highly resistive oxides, the improvement of oxide conduction by surface engineering (metal doping for instance) would be an efficient way to make the present procedure applicable to such samples.

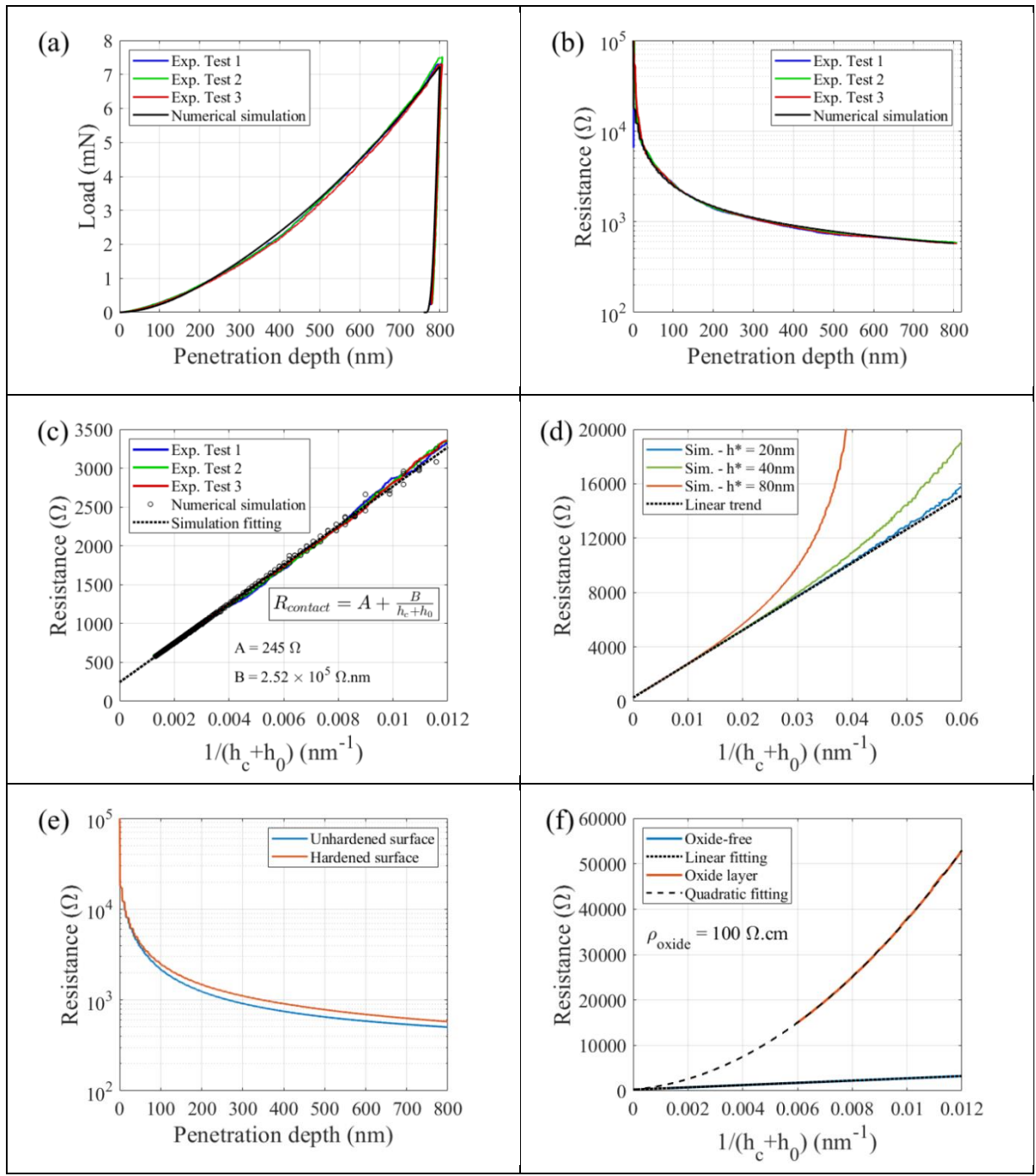

Figure 10: Results of modelled resistive-nanoindentation tests and comparison to experimental results. (a) Mechanical load-depth curves (b) Resistive-nanoindentation curves. (c) Resistance vs the reciprocal of the 'corrected contact depth'. (d) Validity domain of Eq. 17. (e) Effect of rheology on the contact resistance. (f) Effect of an oxide layer.

4.3.2.2. Current line distribution across the tip-sample interface

As already discussed in Section 3.2, the linear dependence of resistance with the reciprocal of contact radius relies on the assumption that the shape factor $SF$ remains constant during indentation [53]. However Nakamura et al focused on a highly-symmetrical system: both sides of the contact are semi-infinite solids with identical resistivities. The present system is drastically different: while sample is a highly-conducting semi-infinite solid, tip is a highly-resistive self-similar pyramid. Consequently, this assumption has to be verified numerically for this resistive-nanoindentation system.

Similarly to the framework proposed in [53], the current line distribution across the interface is calculated against the size of the peripheral ring normalized to the total contact radius (see Fig. 6 b and [53] for details). First the constriction of current lines at the contact periphery is clearly shown in Fig. 11 a. Fig. 11 b reports the shape factor $SF$ for different penetration depths for the Berkovich tip. This figure clearly shows that all these curves superimpose, thus confirming the 'constant $SF$' assumption. It also shows that ~90% of the current crosses a virtual ring of half the radius. For this ring dimension, the relative difference lies below 2 % for penetration depth varying from 15 nm to 1000 nm.

In order to extend our method to the use of sharper tips than Berkovich geometries, the case of cube-corner tips is also considered. The shape factors for Berkovich and cube-corner tips are thus compared to the reference case (highly-symmetrical system) in Fig. 11 c. It is clear that the worst case is found with Berkovich geometry, where the combination of highly-resistive tip to large cone-angle forces a stronger current constriction at the contact periphery. As the 'constant $SF$' assumption has already been validated for the Berkovich geometry, the method can be safely extended to cube-corner geometry that leads to a shape factor $SF$ curve closer to the reference case.

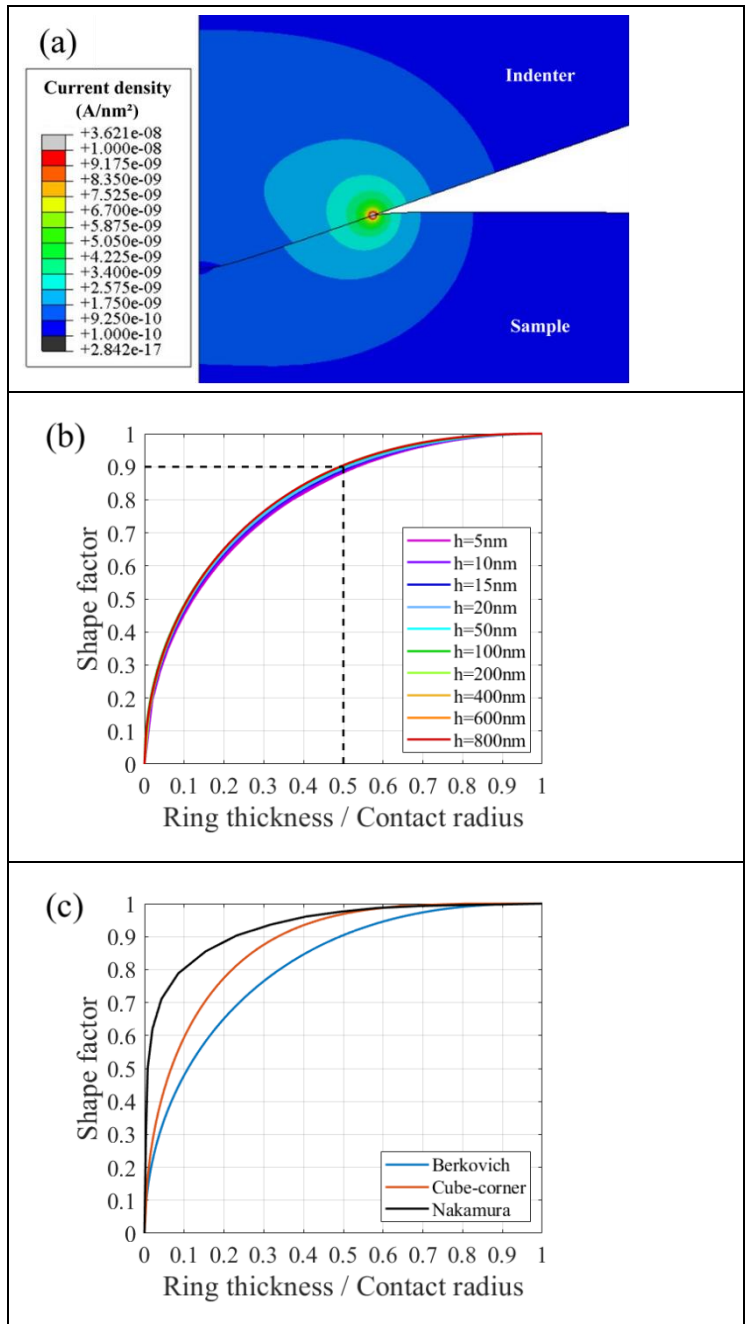

Figure 11: Construction of current line at contact periphery. (a) Distribution of current density. (b) Shape factors with a Berkovich tip for different penetration depths. (c) Shape factors with different tip geometries for a penetration depth of 800 nm.

## 4.4. Data processing and numerical modeling of resistive-nanoindentation tests on Au thin films

4.4.1. Analytical data processing: contact area monitoring

Fig. 12 a reports a set of resistive-nanoindentation tests performed on a Au thin film deposited on sapphire (film thickness $t = 200$ nm). This structure has been chosen for its complex rheology: the film thickness adds an extra length scale to the system, delamination may occur at the interface,… The standard practice to analyze such a soft-film-on-stiff-substrate structure is usually to determine the contact area by numerically solving a Fredholm integral [60-62] (or using finite element simulations) coupled with at least one post-mortem AFM imaging of indent imprints at a given depth. More empirical approaches were also developed [63,64].

Various final penetration depths have been tested (50, 100 and 150 nm, several tests for each). Thanks to the calibration step that relates contact resistance to contact depth (Eq. 18), it is possible to extract the contact area continuously during the tests. Fig. 12 b compares the contact areas extracted from this procedure to AFM data. Once more excellent agreement is found: standard deviation as low as 4% in average is obtained. It has to be noted that a strong advantage of resistive-nanoindentation is the ability to measure contact area continuously during the test. Only the use of BDD tips allows such a calibration-based area monitoring procedure by cancelling its dependence to sample resistance $R_{Sample}$.

For the sake of comparison, contact areas were also calculated with two analytical methods proposed in literature: the Oliver-Pharr [33] and Saha-Nix methods [64]. The former is the most widely used method in practice for nanoindentation analysis, the second is a semi-empirical method dedicated to thin film characterization. The corresponding results are reported in Fig. 12 b. It can be seen that these two methods underestimate the actual contact area: Oliver-Pharr method underestimates the contact area all along indentation by ~50%. The Saha-Nix method describes correctly the system for low penetration depths (up to ~40% of the film thickness) but deviates for deeper indents, as discussed by Han et al [60].

4.4.2. Numerical modeling

The numerical modeling of these experiments was also performed. The tip characteristics were kept rigorously identical to the bulk Au case. As far as the sample is concerned, its film-on-substrate structure has been modelled by considering a fully tied interface. The constitutive law for Au was adjusted to fit the loading curves (Fig. 12 c). The resulting evolution of the contact area against depth is also reported in Fig. 12 b. This plot appears to fit reasonably to the experimental data, even though the curve inflexion is not reproduced (weaker inflexion for the numerical plot). This later point suggests that some extra behaviors should be considered for a better fit (e.g. more complex constitutive law of the film, possible interface delamination, or intrinsic film stress). In addition, a full-3D approach could be more appropriate to model this system. It is to be noted that a better description of the system is necessary to improve numerical modeling but it would be pointless for the analysis of experimental data. This latter point illustrates the ability of resistive-nanoindentation to supply extra inputs for the quantitative analysis of nanoindentation.

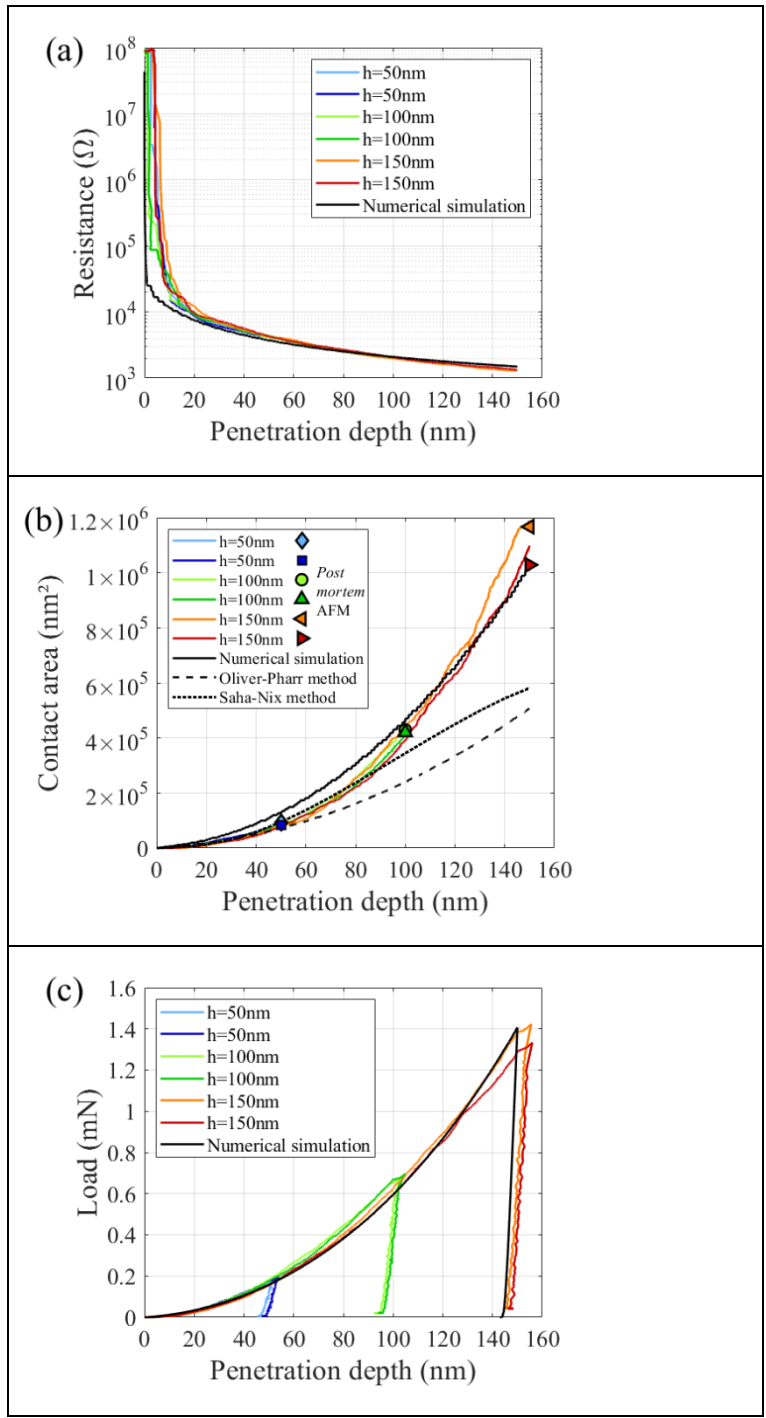

Figure 12: Results of modelled resistive-nanoindentation tests performed on Au thin film and comparison to the experimental data. (a) Resistive-nanoindentation curves. (b) Evolution of contact area against penetration depth (resistive-nanoindentation data, AFM observations, literature models and present numerical model). (c) Mechanical load-depth curves.

# 5. Conclusion

The present work reports the experimental, analytical and numerical description of resistive-nanoindentation tests performed on gold (bulk and thin film). It is first shown that the electrical contact resistance can be fully described throughout nanoindentation timeline as a succession of tunneling, ballistic and diffusive conduction mechanisms. Then a numerical framework is used to assess the technique capabilities (validity domains, sensitivity to tip defect, sensitivity to rheology, effect of an oxide layer,…) and to validate assumptions made on current line distribution within the system. Finally, the technique has shown its strength for the monitoring of contact area on samples with complex rheology. The application of this technique should now be extended to other fields. First the characterization of natively-oxidized metals is of primary interest for general applications. The ability of resistive-nanoindentation to sense the cracking of the oxide layer (inducing resistance drops) or electrochemical processes (conductivity change after oxide reduction) could be used for the fine characterization of such mechanisms. Another application field would be to access the sample resistivity simultaneously to its mechanical properties. For now, the doped diamond tip resistance overwhelms the sample one, thus making the technique insensitive to material resistivity: a decrease of tip resistivity would bring the sample contribution in the measurable range. These points strongly highlight the versatility and flexibility of this technique, as well as its potential for further developments.

# 6. Acknowledgments

This work has been performed with the financial support of the Centre of Excellence of Multifunctional Architectured Materials "CEMAM" n° ANR-10-LABX-44-01. The CEMAM program is funded by the French Agence Nationale de la Recherche (ANR).

The authors thank the technical team of SIMaP lab (Bruno Mallery, Stéphane Massucci and Nadine Vidal) for its support, as well as members of CSI/Scientec company (Les Ulis, France) for their support for the functionalization of the nanoindentation head: Louis Pacheco, Alan Lecoguiec and Sylvain Poulet.

The authors would also like to sincerely thank E. Siebert and J. Fouletier from LEPMI lab for fruitful discussions.

# 7. Appendix: Details on finite-element modeling

Calculations were carried out using the Finite Elements Method (FEM), with the ABAQUS software [47]. The mechanical model is set up in a 2D axisymmetric framework. It consists of two separated parts, further denoted 'the indenter' and 'the sample' respectively (Fig. 13):

- An axisymmetric indenter is considered for the calculations, representing the Berkovich indenter that was used for the experiments. The 'shape function' $A_c(h_c)$ of the indenter was the one measured experimentally (with $h_c$ taken from the tip apex). Although the modeled indenter and the real indenter do not have the same general shape, the radius of the modeled indenter is chosen so that the two 'shape functions' match (i.e. the modeled indenter radius is $r_c(h_c) = \sqrt{\frac{A_c(h_c)}{\pi}}$).

- The sample is defined as an elastic-plastic cylinder with a 10μm radius and height.

A rigid frictionless contact is defined between the indenter surface and the sample top surface.

The general idea of the simulation is to *simultaneously* compute the deformation of the bodies (indenter and sample) and the distribution of currents inside the system. For this purpose, a coupled temperature-displacement procedure is used. The rationale for this choice of procedure is now explained: it is not our purpose to model heat exchange in this work, but we use the temperature field to mimic the electric potential field, thereby assuming that the electrical conduction is following a pure Ohmic law. Keeping this in mind, we will talk about potential and currents in the following of the text.

Because of the strong deformations typical from indentation problems, the calculations are carried out in the framework of large displacements. The solver used is ABAQUS standard, so the calculation is quasi-static (equilibrium is reached at the end of each calculation increment).

Let us now describe the material properties that have been chosen. They are all displayed in Table 2. The indenter is modeled as an isotropic elastic linear body, with Young's modulus $E_i$ and Poisson's ratio $v_i$.

The 'Bulk sample' (Fig. 13 a) is modeled as an elastic-plastic material. It is assumed to be isotropic linear in the elastic regime, with Young's modulus and Poisson's ratio respectively denoted $E_s$ and $v_s$. In plastic regime, an isotropic hardening is assumed, following a Hollomon power law:

$$\sigma_y = \sigma_{y0,b} + K\varepsilon_{p,eq}^n \qquad (A1)$$

where $\sigma_y$ is the yield stress, $\varepsilon_{p,eq}$ the equivalent plastic strain, $\sigma_{y0,b}$ the initial yield stress, $K$ and $n$ two constants. $\sigma_{y0,b}$, $K$ and $n$ were used as fitting parameters in order to reproduce the experimental Load-Depth curves (see Tab. A1 for values).

The top-surface hardening induced by polishing was taken into account by a specific hardening law implemented on a 900 nm-thick top layer of the sample (Fig. 13 a). An isotropic linear hardening law has been chosen:

$$\sigma_y = \sigma_{y0,l} + E_T\varepsilon_{p,eq} \qquad (A2)$$

where $\sigma_y$ is the yield stress, $\varepsilon_{p,eq}$ the equivalent plastic strain, $\sigma_{y0,l}$ the initial yield stress and $E_T$ the tangent modulus (see Tab. A1 for values).

| Part | Indenter | Bulk sample | Thin film sample | Thin film sample | Unit |
|------|----------|-------------|------------------|------------------|------|
| Material | BDD | Au | Au film | Sapphire substrate | |
| Elastic behavior | $E_i = 1129$ | $E_s = 76$ | $E_s = 76$ | $E_{ss} = 400$ | GPa |
| | $\upsilon_i = 0.07$ | $\upsilon_s = 0.42$ | $\upsilon_s = 0.42$ | $\upsilon_{ss} = 0.29$ | |
| Plastic behavior (Au bulk or film) | | $\sigma_{y0,b} = 60$ | $\sigma_{y0,b} = 150$ | | MPa |
| | | $k = 40$ | $k = 410$ | | MPa |
| | | $n = 0.1$ | $n = 0.1$ | | |
| Plastic behavior (hardened Au layer) | | $\sigma_{y0,l} = 100$ | | | MPa |
| | | $E_T = 1.8$ | | | GPa |
| Electrical behavior | $\rho_i = 2.2 \cdot 10^{-1}$ | $\rho_s = 2.2 \times 10^{-6}$ | $\rho_s = 2.2 \times 10^{-6}$ | | $\Omega.cm$ |

Table 2: Materials properties used in FEM simulations.

The properties of the 'Thin film sample' (Fig. 13 b) are also given in Table 2. The 200 nm-thick Au film is modeled as an elastic-plastic material, while the sapphire substrate is considered as purely linear elastic. The film is fully tied to the substrate (i.e. nodes are shared by the film and substrate elements along the film/substrate boundary). In order to minimize the influence of the system boundaries, both substrate height and radius were taken to be 100 times larger than the film thickness.

Concerning the electrical properties, electrical conduction in the indenter and in the sample was modeled using a pure ohmic law with an electrical conductivity extracted from previous works [65] (Tab. 1). The resistivities of Au either in the 'Bulk', 'hardened surface' or 'thin film' parts were kept identical (mechanical hardening is expected to modify weakly Au resistivity [66]). An electric contact interaction property has been defined between the indenter tip and the sample surface (in addition to the mechanical contact property previously mentioned). It consists of a conductance between the two surfaces defined as a function of the gap $d$ between the surfaces (node to node distances along the directions normal to the surfaces, also called 'clearance'). In our case, the conductance is $0\ \Omega^{-1}.\text{cm}^{-2}$ for $d \geq 1$ nm, and $10^{+11}\ \Omega^{-1}.\text{cm}^{-2}$ ('infinite conductivity') for $d = 0$ nm (with linear interpolation for intermediate values, $0\ nm \leq d \leq 1\ nm$).

We now describe the loading and boundary conditions. The general idea of the loading is to prescribe a displacement at the top of the indenter part in order to make the tip penetrate into the sample, while a potential bias is prescribed between the top of the tip and the bottom of the sample. This loading is actually divided in 3 steps:

- Step1: Setting of the electrical bias. A $V_0$ bias is applied between the top surface of the tip (0 V) and the bottom surface of the sample ($V_0$). Simultaneously, a displacement of very small amplitude (1 nm) is applied at the top of the indenter part, in order slightly increase the initial contact area (i.e. in order to get a finite size contact area instead of a point). This is intended to improve the convergence of this first step of electrical calculation.

- Step2: Indentation - loading stage. A downward displacement of 800 nm is applied to the top of the indenter part, while the $V_0$ bias is maintained. Because of the non linearities of the problem (involving both contact and plasticity), small increments of displacements are progressively applied, up to the total target displacement of 800 nm. The steady state electrical conduction problem is solved at each increment.

- Step3: Indentation - unloading stage. The indenter is displaced upwards, until the contact between the tip and the sample vanishes. The procedure is the same as the one used in step 2, the $V_0$ bias is also maintained during this calculation step.

The two parts are meshed with 3-nodes axisymmetric linear interpolation for displacement and temperature elements (CAX3T in ABAQUS). A mesh bias was used, with smaller elements in the indenter/sample contact area (1 nm and 4 nm large for the 'Bulk sample' and the 'Thin film sample', respectively).

As part of the post-processing, the contact radius was extracted by exploiting the contact pressure data along the substrate top surface. Using a Python program, the elements of the top surface of the substrate were explored one by one for increasing distances from the symmetry axis. The distance for which the contact pressure vanishes has been taken to be the contact radius. Additionally, the electrical resistance has been determined by measuring the current flux crossing the surface of the indenter[1]. Finally, the reaction force of the substrate on the indenter was measured directly by summing the vertical nodal reaction forces components of the top surface of the indenter (variable RF2 in ABAQUS).

---

[1] In our context of a thermal-mechanical analysis, it is rather the heat flux which is processed (variable HFLA in ABAQUS).

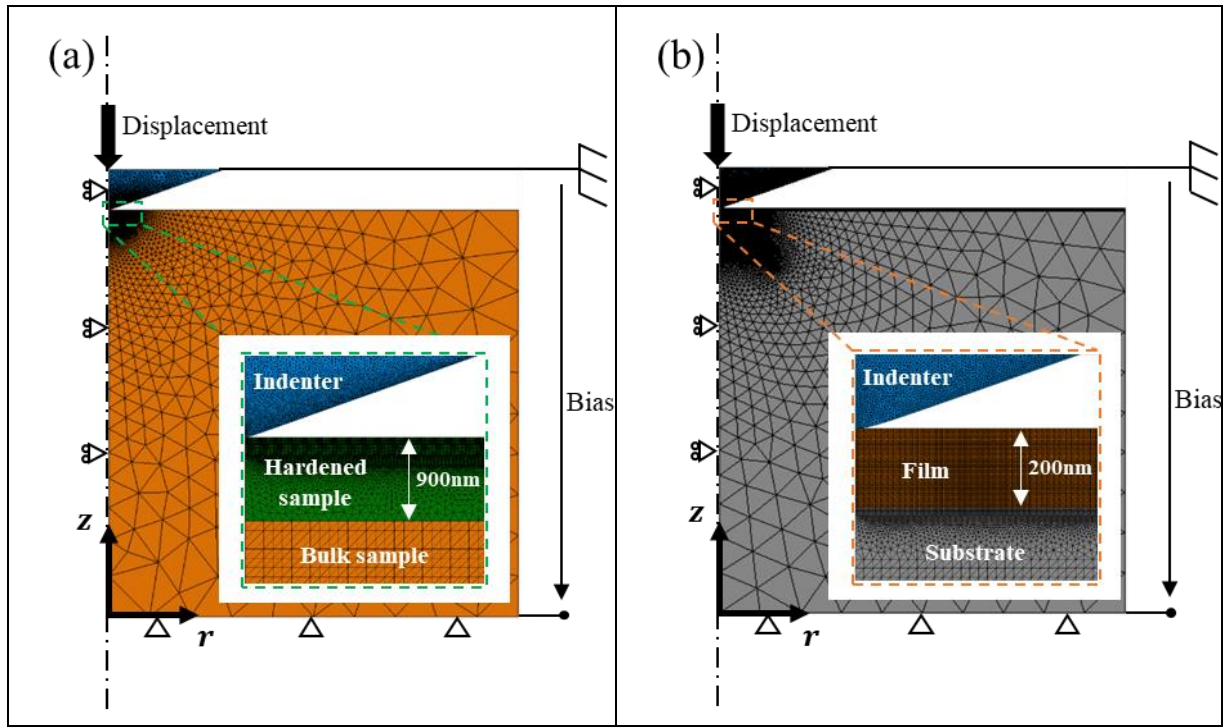

Figure 13: Modeled system. (a) Bulk sample including the hardened surface. (b) Thin film sample.

# 8. Data Availability Statement

The data that supports the findings of this study are available within the article.